\documentclass[prb,showpacs,preprintnumbers,amsmath,amssymb,letterpaper,twocolumn]{revtex4}

\usepackage{graphicx}
\usepackage{dcolumn}
\usepackage{bm}
\usepackage{epstopdf}
\usepackage{latexsym}
\usepackage{epsfig}
\usepackage{verbatim}
\usepackage{hyperref}

\newcommand{\be}{\begin{equation}}
\newcommand{\ee}{\end{equation}}
\newcommand{\bea}{\begin{eqnarray}}
\newcommand{\eea}{\end{eqnarray}}

\renewcommand{\Re}{\mathrm{Re}}

\newcommand{\eref}[1]{Eq.~(\ref{#1})}
\newcommand{\reqs}[1]{Eqs.~(\ref{#1})}
\newcommand{\rref}[1]{(\ref{#1})}

\newcommand{\ocite}[1]{Ref.~\onlinecite{#1}}

\begin{document}

\title{Pairing resonance as a normal-state spin probe in ultra-thin Al films}

\author{G. Catelani}
\affiliation{Department of Physics and Astronomy, Rutgers University, Piscataway, New Jersey 08854, USA}
\author{Y. M. Xiong, X. S. Wu}
\altaffiliation[Present address:]{School of Physics, Georgia Institute of Technology, Atlanta,
Georgia 30332, USA}
\author{P. W. Adams}
\affiliation{Department of Physics and Astronomy, Louisiana State University, Baton Rouge,
Louisiana 70803, USA}

\date{\today}

\begin{abstract}
We present a quantitative analysis of the low-temperature,
high parallel field pairing resonance in
ultra-thin superconducting Al films with dimensionless conductance
$g\gg 1$.  In this regime we derive an analytical expression for the
tunneling density-of-states spectrum from which a variety of
normal-state spin parameters can be extracted.  We show that by
fitting tunneling data at several supercritical parallel magnetic
fields we can determine all of the relevant parameters that have
traditionally been obtained via fits to tunneling data in the
superconducting phase. These include the spin-orbit scattering rate,
the antisymmetric Landau parameter $G^0$, and the orbital
pair-breaking parameter.

\end{abstract}


\maketitle

\section{Introduction}
\label{sec:intro}

Determining the microscopic spin parameters of paramagnetic metals
has historically been a process fraught with complications and
inaccuracies.\cite{Gibson1989,Vier1984}  In general, the spin
response of an interacting fermionic system can be modified by
spin-orbit scattering processes, electron-phonon interactions,
and/or electron-electron interactions.\cite{Baym1991,Leggett1965}
These contributions to the spin susceptibility themselves can be
affected by disorder,\cite{Altshuler1987,Bergmann1985}
dimensionality,\cite{Mackenzie2003,Aleiner2007} and the presence of
interfaces.\cite{Gorkov2001} The two primary spin parameters for a
paramagnetic system are the spin-orbit scattering rate and the
antisymmetric $l=0$ Landau parameter $G^0$.  The latter accounts for
the renormalization of the bare Pauli spin susceptibility due to
electron-phonon and electron-electron interactions.  Depending upon
the sign of this parameter the effective spin moment can be larger
or smaller than the bare electron value.   In practice, the
spin-orbit scattering rate can be obtained from the coherent
backscattering contributions to the magneto-resistance of moderately
disordered non-superconducting films or by parallel magnetic field
studies of thin superconducting films. The Fermi-liquid parameter
$G^0$, however, is more difficult to determine accurately. In
principle, it can be extracted from low-temperature measurements of
the spin susceptibility $\chi$ and the heat capacity $\gamma$ from
which the respective corresponding density of states $N(\chi)$ and
$N(\gamma)$ are obtained. The ratio of these densities of states is
a direct measure of the many body renormalization,
$G^0=N(\gamma)/N(\chi)-1$.\cite{Baym1991}  Unfortunately, orbital
contributions to the susceptibility make it very difficult to
determine its spin component precisely in high-conductivity systems
and phonon contributions to the specific heat can introduce
significant systematic errors in the measurement of $N(\gamma)$. In
this report we address the determination of $G^0$ and the spin-orbit
scattering rate via the Pauli-limited, normal-state pairing
resonance.\cite{Wu1996,Aleiner1997,Butko1999,Adams2000}

If a paramagnetic system has a superconducting phase and can be made
into a thin-film form, then it is possible to access the spin
parameters through tunneling density-of-states (DOS) measurements.  A Zeeman splitting
of the BCS coherence peaks can be induced by applying a parallel magnetic field to a film of
thickness $t\ll\xi$, where $\xi$ is the superconducting coherence length. Tedrow
and Meservey pioneered the use of superconducting spin-resolved
tunneling to directly measure both spin-orbit scattering rate and
the Landau parameter $G^0$ in thin Al and Ga films near the parallel
critical-field transition.\cite{Tedrow1984,Alexander1985,Gibson1989}
This technique, however, cannot access
$G^0$ well into the superconducting phase since those electrons
responsible for the exchange effects are consumed by the formation
of the condensate.\cite{CWA}  To circumvent this limitation, one needs to
measure the Zeeman splittings in magnetic fields just below parallel
critical field.  However, one cannot
completely reach the normal-state quasiparticle density in a thin
film while remaining in the superconducting phase, since the spin-paramagnetically limited
parallel critical-field transition is first-order.  Because of this,
one must extrapolate the normal-state value of $G^0$ from data taken
in the superconducting phase.  Alternatively, the films can be made
marginally thicker, which will suppress the first-order
transition,\cite{CWA} or the measurements can be made at higher temperatures.
But these strategies limit one to a very narrow range of film
thicknesses.  Furthermore, in both cases one is constrained to a
very narrow range of applied fields.

Here we present a detailed analysis of the normal-state pairing
resonance (PR) from which the spin-orbit scattering rate, orbital depairing
parameter, and the Landau parameter $G^0$ can be accurately
obtained.  We show that the technique can be used over a wide range
of film thicknesses and resistances. Moreover, the measurements
can be made in fields well above the parallel critical field and in
fields substantially tilted away from parallel orientation.\cite{WAC,Catelani2006}

\section{Pairing resonance in parallel field}
\label{sec:prpf}

The PR is characterized, as any other resonance, by two quantities:
its position and its width. The former is given by\cite{Aleiner1997}
\be\label{Epdef}
E_+ = \frac{1}{2}\left(E_Z + \Omega \right) \ ,
\ee
where
\be\label{ezdef}
E_Z = \frac{2\mu_B H}{1+G^0}
\ee
is the Zeeman
energy renormalized by the Fermi-liquid parameter $G^0$, $\mu_B$ is
the Bohr magneton, and
\be
\Omega=\sqrt{E_Z^2-\Delta_0^2}
\ee is the
Cooper-pair energy with $\Delta_0$ the zero-field, zero-temperature
gap of the corresponding superconducting phase.

The width of the PR depends on the effective dimensionality of the
sample and on the strength $\Gamma$ of pair-breaking mechanisms
other than the Zeeman splitting. If these are absent, a
non-perturbative approach is necessary (see \ocite{Aleiner1997}), and for quasi-two
dimensional systems the width is
\be
W_2 = \frac{\Delta_0^2}{4g\Omega} \ ,
\ee
where $g = 4\pi\hbar\nu_0 D$ is the
dimensionless conductance with $D$ the diffusion constant and
$\nu_0$ the bare DOS. If $W_2\ll \Gamma$ then a perturbative
calculation is sufficient to accurately estimate the width, provided
one properly takes into account the role of the exclusion
principle.\cite{Catelani2006} For instance, in the case of a tilted
magnetic field, $\Gamma$ is proportional to the perpendicular
component of the field and the exclusion principle both shifts and
reshapes the PR.  If we consider the effects of spin-orbit
scattering and the finite-thickness orbital contributions of the
parallel field,\cite{innote} then
\be\label{Gdef} \frac{\Gamma}{2\Delta_0} = b +
c \left(\frac{\mu_B H}{\Delta_0}\right)^2 \, ,
\ee
where, according to the notation commonly used to characterize
the DOS in the superconducting state,\cite{Fulde1973}
\be
b=\frac{\hbar}{3 \tau_\mathrm{so} \Delta_0}
\ee
is proportional to the spin-orbit scattering rate $1/\tau_\mathrm{so}$ and
\be
c = \frac{De^2t^3 \Delta_0}{8\ell \mu_B^2 \hbar}
\ee
is the orbital de-pairing parameter, where $t$ is the film's thickness, $e$ is the electron charge,
and $\ell$ is the mean-free path.  This latter parameter quantifies the
strength of the orbital effect of the field\cite{cnote} in relation
to the Zeeman effect.  The Zeeman splitting is the dominant pair-breaking mechanism for
$c\lesssim 1$.

Following the procedure outlined in \ocite{Catelani2006}, we obtain the zero-temperature
correction to the (spin-down) DOS due to the PR
\be\label{dos}
\frac{\delta\nu (\epsilon) }{\nu_0} = -A(\epsilon; E_Z, \Gamma)
\frac{W_2 \Gamma}{(\epsilon - E_+)^2+\Gamma^2} \ ,
\ee
where $\epsilon$ is the energy measured from the Fermi level; the other quantities entering this
formula have been defined above, see \reqs{Epdef}-\rref{Gdef}.
The correction for the other spin component is found by replacing $\epsilon \to -\epsilon$ in
the right-hand side of \eref{dos}. The function
\be\label{amp}\begin{split}
A (\epsilon; E_Z, \Gamma) = \frac{1}{\pi} \Big\{ \arctan[(E_Z - \epsilon)/\Gamma]
+ \arctan[\Omega/\Gamma] \ \\ +
\arctan[(\epsilon - \Omega)/\Gamma] + \arctan[(2 \epsilon - E_Z)/\Gamma] \Big\}
\end{split}\ee
accounts for the exclusion principle and takes on values between 0
and 2.   It alters the Lorentzian shape of the PR, especially at
energies close to the Fermi energy (i.e., $\epsilon \ll E_+$) and,
in fact, $A(\epsilon=0)=0$. We note that \reqs{dos}-\rref{amp} imply that
$\delta\nu/\nu_0 \leq 2W_2/\Gamma$, which is consistent with the
assumed perturbative criterion $\Gamma\gg W_2$.

In this work we show that \eref{dos} gives a quantitative description of the PR and that it
enables us to extract from normal-state measurements the physical quantities $G^0$, $b$, and $c$.
While they can be obtained from DOS measurements in the superconducting
state,\cite{Alexander1985,CWA} this requires
to solve a set of self-consistent equations for the order parameter and ``molecular'' magnetic
field together with the Usadel equations for the normal and anomalous Green's functions -- a much
more complicated task in comparison to the simple fitting of the data that we describe
in Sec.~\ref{sec:res}.

\section{Experimental procedure}
\label{sec:ep}

Aluminum films were grown by e-beam deposition of 99.999\% Al stock
onto fire-polished glass microscope slides held at 84~K. The
depositions were made at a rate of $\sim0.1$~nm/s in a typical
vacuum $P<3\times10^{-7}$~Torr. A series of films with thicknesses
ranging from 2 to 2.9~nm had a dimensionless normal-state
conductance that ranged from $g=5.6$ to $230$ at 100 mK. After
deposition, the films were exposed to the atmosphere for 10-30 min
in order to allow a thin native oxide layer to form. Then a 9-nm
thick Al counterelectrode was deposited onto the film with the oxide
serving as the tunneling barrier. The counterelectrode had a
parallel critical field of $\sim$2.7~T due to its relatively large
thickness, which is to be compared with $H_{c\parallel}\sim6$~T for
the films. The junction area was about 1~mm$\times$1~mm, while the
junction resistance ranged from 10-100~k$\Omega$, depending on
exposure time and other factors.  Only junctions with resistances
much greater than that of the films were used.  Measurements of
resistance and tunneling were carried out on an Oxford dilution
refrigerator using a standard ac four-probe technique. Magnetic
fields of up to 9~T were applied using a superconducting solenoid. A
mechanical rotator was employed to orient the sample \textit{in
situ} with a precision of $\sim0.1^{\circ}$.

\section{Results and discussion}
\label{sec:res}

\begin{figure}
\begin{flushleft}
\includegraphics[width=.48\textwidth]{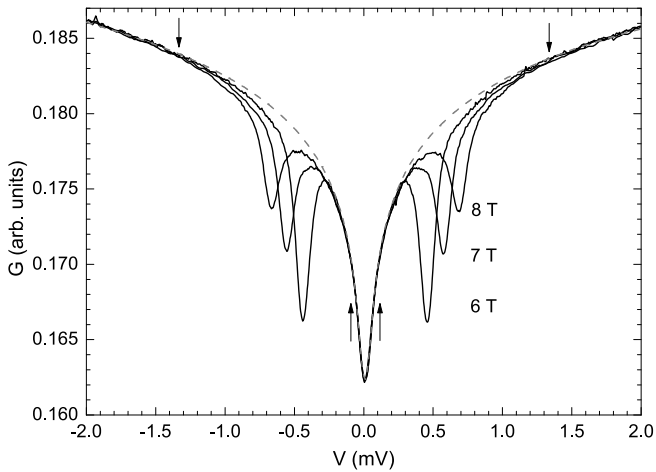}\end{flushleft}
\caption{\label{Data}Tunneling conductance at 70~mK for three supercritical parallel magnetic
fields (solid lines).  The dashed line is the best fit to the zero bias anomaly due to Coulomb
interaction. The arrows point to the boundaries of the low- ($|V|\lesssim 0.2$~mV) and
high-bias ($|V|\gtrsim 1.4$~mV) regions used for the fitting.}
\end{figure}

We show in Fig.~\ref{Data} the tunneling conductance measured at
70~mK and three supercritical parallel magnetic fields.  This
particular film of dimensionless conductance $g\simeq 57$ was 2.6~nm thick and had a zero-field
superconducting transition temperature $T_\mathrm{c} = 2.74$~K.
Common to the three data sets is the Coulomb zero-bias
anomaly (ZBA),\cite{AA85} which produces a logarithmic depletion in the DOS at high
biases; the logarithm is cut off at low bias by temperature. In
order to isolate the paramagnetic resonance, we need to remove the
contribution of the ZBA. To interpolate between the
low- and high-bias parts of the curves (as delimited by the arrows in
Fig.~\ref{Data}), we find the best-fit curve, restricted to these
regions, given by the sum of a background constant tunneling
conductance and $\Re~\Psi (1/2 + i \alpha V)$, where $\Psi$ is the
digamma function and $\alpha$ a fitting parameter. The result is the
dashed curve in Fig.~\ref{Data}, which is then subtracted from the
measured tunneling conductances.

\begin{figure}
\begin{flushleft}
\includegraphics[width=.485\textwidth]{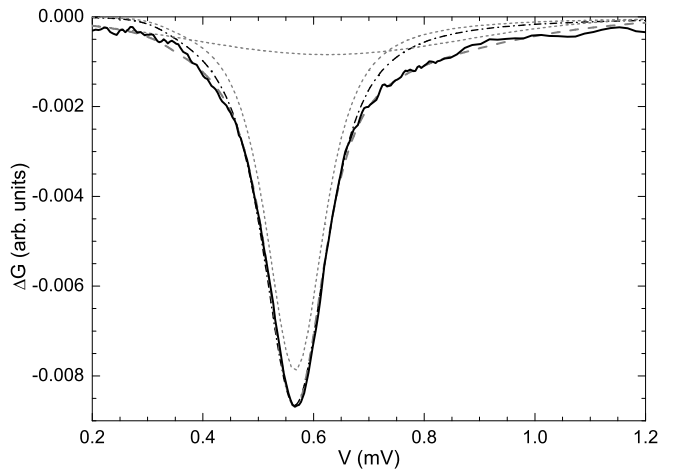}\end{flushleft}
\caption{\label{Fit7}Pairing resonance at 7~T (solid line) with the ZBA subtracted off. The dot-dashed
curve is the best fit to the data using \eref{dos}. The dashed curve
is the best fit with a sum of \eref{dos} and a Gaussian -- see the text for more details on the
fitting procedure. The two terms of the sum are plotted separately as dotted curves.}
\end{figure}

In Fig.~\ref{Fit7} we plot with a solid line the PR at 7~T obtained as described above. As discussed
in Sec.~\ref{sec:prpf}, its position and width are respectively determined by
the Zeeman energy $E_Z$ and the pair-braking rate $\Gamma$, while the conductance $g$ only affects
the overall magnitude. Using \eref{dos},
the best fit to the data is given by the dot-dashed curve;
while the main peak is well reproduced, a shoulder
feature at higher bias is underestimated. To our knowledge,  there are two possible causes
for this discrepancy, namely a
finite bias, triplet channel anomaly,\cite{AA85} similar to the Coulomb ZBA but much weaker,
and finite-temperature effects.\cite{tbrnote} To take into account these possible
corrections, we add to \eref{dos} a Gaussian contribution; to reduce the number of free parameters,
we require it to be centered at the Zeeman energy, which is where a triplet channel correction would
be located, while
the amplitude and width are used as fitting parameters. The best fit thus found is the dashed line in
Fig.~\ref{Fit7}; the peaked PR and broad Gaussian contributions are plotted separately with dotted lines.

\begin{figure}
\includegraphics[width=.465\textwidth]{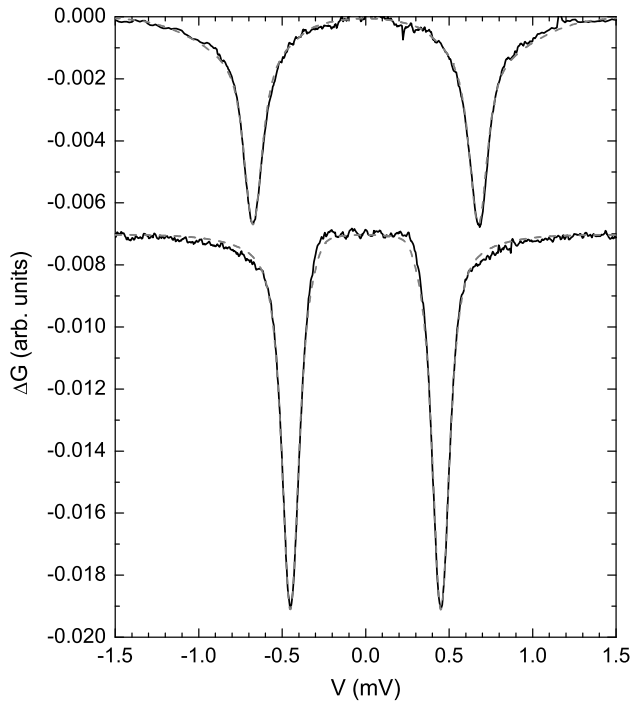}
\caption{\label{Fit68}Pairing resonances measured at 8~T (top) and 6~T (bottom solid curve).
The bottom curve is shifted down by 0.007 for clarity. The dashed lines are best fits to the data obtained
as described in the text. The asymmetry of the PR and
its suppression near the Fermi energy are easily recognized in the data taken at 6~T.}
\end{figure}

We present in Fig.~\ref{Fit68} two more PRs with the best-fit curves.
The asymmetric shape of the resonance and its suppression near the Fermi energy are evident in
the lowest field data. We note that fitting these data with
\eref{dos} only would require us to decrease the conductance with increasing field,
whereas we can use the same value of the conductance at all fields when the Gaussian correction is included.
Moreover, the value of the Zeeman energy is only weakly affected by the inclusion of this correction, with
the change in $E_Z$ smaller than our estimated relative error of about 1\%. While these two
observations support the validity of our approach, the magnitude of the width parameter $\Gamma$
turns out to be more sensitive to the Gaussian correction.
However, its field dependence (see Fig.~\ref{GEz}) is robust, and the
quantitative estimates discussed below are in line with expectations.

\begin{figure}
\begin{flushleft}
\includegraphics[width=.48\textwidth]{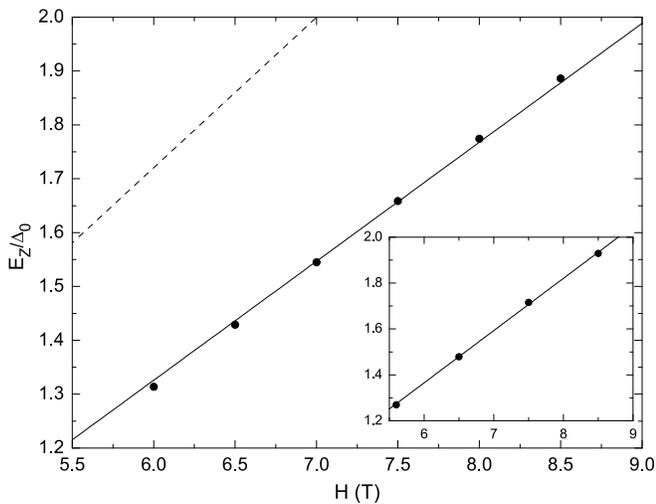}\end{flushleft}
\caption{\label{EzH}Normalized Zeeman energy $E_Z/\Delta_0$ vs. magnetic field $H$. The solid line
is the best fit to \eref{ezdef};
the slope is proportional to $(1+G^0)^{-1}$, and we estimate the value of the Fermi-liquid
parameter $G^0\simeq 0.26$. For comparison, the dashed
line represents the expected linear relationship in the absence of Fermi-liquid renormalization.
Inset: same plot as the main figure, but for a thicker film with $G^0\simeq 0.24$ (see text for details).}
\end{figure}

Having detailed our fitting procedure, we now consider the physical quantities that can be extracted from
the data. In Fig.~\ref{EzH} we plot the normalized Zeeman energy as a function of the applied field.
By fitting the data with \eref{ezdef} we find $G^0\simeq 0.26$; a similar estimate, $G^0\simeq 0.24$,
is obtained for a thicker film with $t=2.9$~nm, $g=230$, and zero-field, zero-temperature
gap $\Delta_0 =0.41$~meV -- see the inset of Fig.~\ref{EzH}.
We note that a better fit to the data in Fig.~\ref{EzH} could be obtained by allowing
for a finite negative intercept;
however, the large estimated error on the intercept makes the best-fit line compatible with the
expectation that it passes through the origin [see \eref{ezdef}].
This finite intercept could be due to small higher-order contributions, since at the lowest field
the parameter $2W_2/\Gamma \simeq 0.07$ is only marginally smaller than 1. In support to this interpretation,
we find no evidence of finite intercept for the thicker film for which $2W_2/\Gamma \lesssim 0.016$.
Alternatively, the intercept could be an additional indication, together with the shoulder feature
mentioned above, of finite-temperature effects. We will further investigate this latter issue in a
separate work.

\begin{figure}
\begin{flushleft}\includegraphics[width=.48\textwidth]{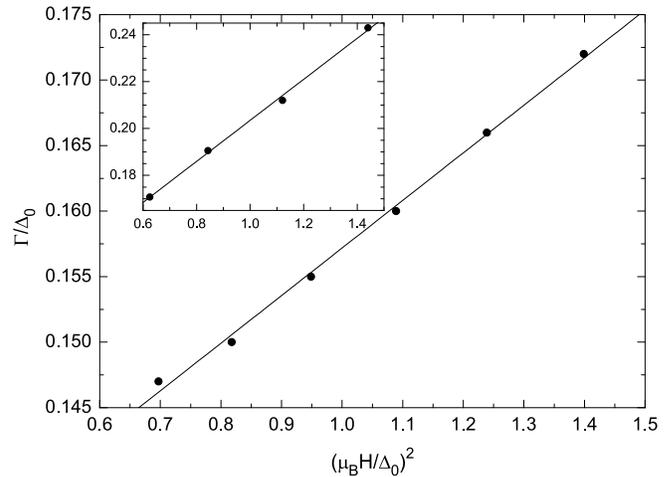}\end{flushleft}
\caption{\label{GEz}Normalized pair-breaking parameter $\Gamma/\Delta_0$ vs. the square of the reduced
field. Using
the linear relationship in \eref{Gdef} we obtain from the best-fit line the spin-orbit scattering rate
$b\simeq 0.06$ and the orbital effect parameter $c\simeq 0.02$. As in Fig.~\ref{EzH}, we show in the inset
the data pertaining to the 2.9~nm-thick film.}
\end{figure}

\begin{figure}
\begin{flushleft}\includegraphics[width=.485\textwidth]{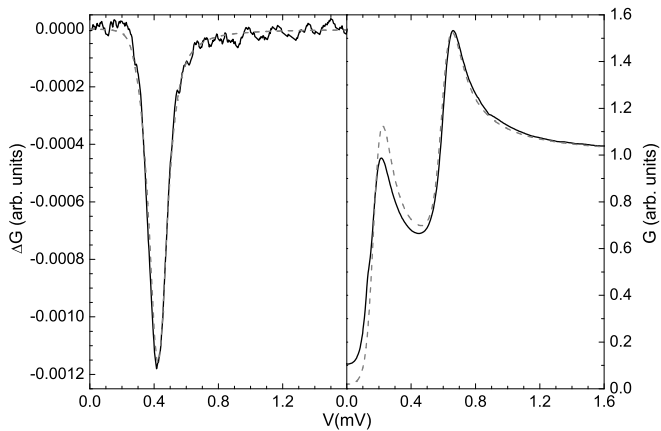}\end{flushleft}
\caption{\label{NSdos}Tunneling DOS in the normal (left, $H=5.6$~T) and superconducting
(right, $H=4$~T) states at $T=70$~mK for a 2.9~nm thick film.
Solid lines are experimental data; dashed lines have been
calculated with the parameters given in the text.}
\end{figure}

The width parameter $\Gamma$ is plotted in Fig.~\ref{GEz} as a function of $(\mu_B H/\Delta_0)^2$
together with
the best-fit line. According to \eref{Gdef}, the intercept and the slope are determined by the spin-orbit parameter $b$ and orbital parameter $c$, respectively. We estimate their values as
$b \simeq 0.06 $, in agreement with the results in the literature, and $c \simeq0.02$, which
favorably compares\cite{ccompnote} with the
value $c\simeq 0.04$ extrapolated from superconducting state measurements in marginally
thick (i.e., $c\simeq 1$) films. Repeating the analysis for the thicker film -- see
the inset of Fig.~\ref{GEz} -- we find $b \simeq 0.06 $ and $c \simeq0.04$. As a further
check on the validity of the present approach, for this film we show
in Fig.~\ref{NSdos} the measured and calculated DOS in the normal and superconducting
states for fields of 5.6 and 4~T, respectively: all the main features of the superconducting DOS are
captured by the theoretical curve\cite{broadnote} obtained by solving the
Usadel and self-consistent equations\cite{Alexander1985} with the
parameters found via the normal-state measurements.

In summary, we have presented a quantitative study of the
paramagnetic pairing resonance in parallel field. We have derived an
expression, \eref{dos}, for the density of states which takes into
account spin-orbit scattering, orbital effect of the magnetic field,
and the Pauli exclusion principle. The latter is responsible for the
suppression of the resonance near the Fermi energy, see
Fig.~\ref{Fit68} and the left panel of Fig.~\ref{NSdos}.
By fitting the PRs measured at different fields we
have obtained the values of the Fermi-liquid parameter $G^0$, the
spin-orbit scattering rate $b$, and the orbital parameter $c$, thus
showing that normal-state experiments can provide the same
information usually extracted from the DOS of the superconducting
phase. Since the PR affects the spin-resolved DOS at opposite biases,
it can, in fact, be used
to probe the electron-spin polarization in magnetic films.  The
present work provides the foundation for the
analysis of tunneling studies of itinerant magnetic
systems via the PR.\cite{PRims}

\acknowledgments

We gratefully acknowledge enlightening discussions with Ilya Vekhter
and Dan Sheehy. This work was supported by the DOE under Grant No.\
DE-FG02-07ER46420 for the experimental portion and by NSF Grant No.
NSF-DMR-0547769 (GC).


\begin{thebibliography}{30}
\bibitem{cnote}This definition is applicable in the limit $t\ll \ell$ of interest here.
\bibitem{tbrnote}We note that the small finite-temperature broadening due to the usual convolution
with the derivative of the Fermi-Dirac distribution function has been taken into account.
\bibitem{ccompnote}The actual film thickness is less than the nominal one due to the oxidation (see
Sec.~\ref{sec:ep}), and extrapolation from thicker films overestimates $c$.
\bibitem{broadnote}In calculating the superconducting DOS we have introduced a small broadening of the
same order of magnitude used in \ocite{CWA}.
\end{thebibliography}
\end{document}